\documentclass[10pt,english,american,preprint, aps, nofootinbib]{revtex4-1}
\usepackage[T1]{fontenc}
\usepackage[latin9]{inputenc}
\usepackage{babel}
\usepackage{units}
\usepackage{mathrsfs}
\usepackage{amsmath}
\usepackage{amssymb}
\usepackage[pdfusetitle,
 bookmarks=true,bookmarksnumbered=false,bookmarksopen=false,
 breaklinks=false,pdfborder={0 0 1},backref=false,colorlinks=false]
 {hyperref}

\makeatletter

\usepackage{etoolbox}
\usepackage{breqn}

\makeatletter
\let\cat@comma@active\@empty
\makeatother

\newcommand{\breqnoverloadothers}
{%
    \renewenvironment{equation}{\ignorespaces\begin{dmath}}{\end{dmath}\ignorespacesafterend}%
    \renewenvironment{equation*}{\ignorespaces\begin{dmath*}}{\end{dmath*}\ignorespacesafterend}%
    \renewenvironment{multline}{\ignorespaces\begin{dmath}}{\end{dmath}\ignorespacesafterend}%
    \renewenvironment{multline*}{\ignorespaces\begin{dmath*}}{\end{dmath*}\ignorespacesafterend}%

}

\newcommand\breqnundefineothers
{%
    \renewenvironment{equation}{}{}%
    \renewenvironment{equation*}{}{}%
    \renewenvironment{multline*}{}{}%

}

\AtBeginEnvironment{dmath}{\breqnundefineothers}
\AtBeginEnvironment{dmath*}{\breqnundefineothers}

\AtBeginEnvironment{dgroup}{\def\breqnundefineothers{}\breqnoverloadothers}
\AtBeginEnvironment{dgroup*}{\def\breqnundefineothers{}\breqnoverloadothers}

\newcommand\brwrap[3]{%
  \setbox0=\hbox{$#2$}
  \left#1\vbox to \the\ht0{\hbox to 0pt{}}\right.\kern-.2em
  \begingroup #2\endgroup\kern-.15em
  \left.\vbox to \the\ht0{\hbox to 0pt{}}\right#3
}

\makeatother

\begin{document}
\title{Effective Field Theory Description of Hawking Radiation}
\author{David A. Lowe}
\affiliation{Department of Physics, Brown University, Providence, RI 02912, USA}
\author{L\'arus Thorlacius}
\affiliation{Science Institute, University of Iceland, Dunhaga 3, 107 Reykjavik,
Iceland\foreignlanguage{english}{ }\\{\rm and}\\ \foreignlanguage{english}{Institutionen
f\"or fysik och astronomi, Uppsala Universitet, \\ Box 803, SE-751
08 Uppsala, Sweden}}
\begin{abstract}
A study is made of Hawking radiation from four-dimensional black holes
using effective field theory methods. The trace anomaly for the stress
tensor in a general curved spacetime background is reproduced using
Riegert's action. The semiclassical stress tensor is evaluated in
a Schwarzschild background taking a time-independent limit for the
quantum state. Imposing physical boundary conditions on an initial
Cauchy surface leads to a unique state, analogous to the Unruh state
with a vanishing ingoing flux and a finite outgoing flux. In particular,
there is no sign of quantum hair arising from this nonlocal effective
action.
\end{abstract}
\maketitle

\section{Introduction}

Hawking's discovery of black hole radiation \citep{HAWKING_1974}
was a triumph of semiclassical gravity and the calculation of Hawking
emission, involving quantized matter fields in a background black
hole geometry governed by the classical equations of general relativity,
is by now textbook material \citep{Birrell:1982ix,Wald:1995yp}. Yet,
fifty years on, it still remains a theoretical challenge to calculate
the semiclassical back-reaction on the black hole metric, due to the
quantum mechanical Hawking emission, in a self-consistent manner in
3+1 spacetime dimensions. In a recent paper \citep{Lowe:2022tun},
we have made progress on the back-reaction problem by formulating
a complete set of semiclassical field equations for 3+1-dimensional
gravity as an initial value problem. The field equations are rather
unwieldy, however, and inherently non-local due to subtleties in regularizing
the quantum stress tensor, which means that physically interesting
solutions are at best amenable to numerical simulation.

Much more progress has been made in understanding the semiclassical
dynamics of black hole radiation in two-dimensional toy models \citep{Teitelboim:1983ux,Jackiw:1984je,Callan:1992rs,Russo:1992ax}.
In particular, the RST model \citep{Russo:1992ax} (a modification
of the earlier CGHS model \citep{Callan:1992rs}) is a two-dimensional
dilaton gravity theory where general time-dependent semiclassical
solutions, with the back-reaction to Hawking emission included, can
be studied in closed form. A key step in solving the model is to adopt
a conformal gauge that renders the non-local Polyakov action in a
local form \citep{POLYAKOV1981207,Callan:1992rs}. 

In the present paper, we explore an effective field theory approach
to black hole radiation in 3+1 spacetime dimensions, somewhat similar
in spirit to the two-dimensional CGHS and RST models. In the higher-dimensional
case, there is no direct analog of the two-dimensional conformal gauge.
However the effective action can be made local by coupling an auxiliary
scalar field to the gravitational background in such way as to reproduce
the original non-local effective action upon integrating out the scalar.\footnote{The analogous construction in two dimensions replaces the Polyakov
term by a scalar field with a standard kinetic term and linearly coupled
to the background Ricci scalar.} The resulting local scalar-tensor theory, which is considerably simpler
to work with than the more general approach of \citep{Lowe:2022tun},
was originally written down some time ago by Riegert \citep{RIEGERT198456},
and has been studied in the black hole context in \citep{Buchbinder:1992gdx,Shapiro:1994ww,Mukhanov:1994ax,Balbinot:1998yh,Balbinot:1999ri,Mottola:2006ew,Anderson:2007aa,Mottola:2016aa,Mottola:2025fhl}.
The essential new result presented here, is a closed form expression
for the effective stress tensor in a Schwarzschild black hole background.
Imposing physical boundary conditions uniquely determines the energy
flux of black hole radiation, with no analog of scalar (or higher-derivative)
hair. This unique configuration is analogous to an Unruh state, \emph{i.e.}
a state that yields a time-independent stress tensor, with vanishing
ingoing flux at past null infinity and finite outgoing flux at future
null infinity. All other solutions violate the semiclassical approximation
either on the future horizon or near past null infinity.\footnote{We note even the present solution violates the semiclassical approximation
at future null infinity, as is well-known for the Unruh state. This
issue is presumably remedied when the semiclassical back-reaction
on the geometry is included and the black hole evaporates.}

Our considerations in this paper are restricted to evaluating a time-independent
semiclassical stress tensor in a static classical black hole background.
Presumably, the semiclassical field equations of the model also have
time-dependent solutions that capture the back-reaction to Hawking
radiation in 3+1 dimensions, but we leave their study for future work.

\section{Basic Setup}

We take as our starting point the following general form of the trace
anomaly in the stress tensor in a curved 3+1-dimensional spacetime,\footnote{In the following, we use Misner, Thorne and Wheeler's so-called $(+++)$
conventions \citep{misner}. Riegert's original paper uses Birrell
and Davies conventions \citep{Birrell:1982ix}, which result in various
sign flips (including, for instance, in the $c\nabla^{2}R$ term in
\eqref{eq:traceanom}).}
\begin{equation}
g^{ab}\left\langle T_{ab}\right\rangle =\frac{1}{16\pi^{2}}\left(aC^{2}+bE-c\nabla^{2}R+dR^{2}+eF^{2}\right)\,,\label{eq:traceanom}
\end{equation}
where
\begin{align}
C^{2} & =R^{abcd}R_{abcd}-2R^{ab}R_{ab}+\frac{1}{3}R^{2},\label{eq:c2e}\\
E & =R^{abcd}R_{abcd}-4R^{ab}R_{ab}+R^{2},\nonumber 
\end{align}
are the square of the Weyl tensor and the Euler density, respectively.
The coefficients $a,b,c,d,e$ depend on the matter content of the
theory. We take the non-vanishing coefficients to be of order $N\gg1$
(while scaling $\hbar\sim1/N$) so that the matter field contribution
to the effective action is dominant compared to that of the metric
sector. We do not consider electromagnetic fields in this work and
hence set $e=0$ in \eqref{eq:c2e}. Furthermore, we follow Riegert
\citep{RIEGERT198456} by imposing a condition on our collection of
matter fields that $d=0$. Duff has shown this condition holds for
ordinary conformally coupled matter fields with spin $\leq1$ \citep{Duff:1993wm}.
With these conditions, it is possible to reproduce the rest of the
trace anomaly via a local tensor-scalar theory \citep{RIEGERT198456,Mottola:2016aa},

\begin{dmath}
\begin{equation}
S=\int d^{4}x\,(-g)^{1/2}\left[\frac{1}{16\pi}R+\frac{1}{192\pi^{2}}(c-\tfrac{2}{3}b)R^{2}-\tfrac{b}{2}\nabla^{2}\phi\nabla^{2}\phi-\tfrac{b}{3}R(\nabla\phi)^{2}+bR^{ab}\nabla_{a}\phi\nabla_{b}\phi+\frac{\phi}{8\pi}\left((a+b)C^{2}+\frac{2b}{3}\left(R^{2}-3R_{ab}R^{ab}-\nabla^{2}R\right)\right)\right].\label{eq:action}
\end{equation}
\end{dmath}The metric equation of motion is given by a somewhat lengthy
expression,

\begin{dmath}
\begin{equation}
0=\frac{G_{ab}}{16\pi}+\frac{(c-\frac{2}{3}b)}{96\pi^{2}}\left((R_{ab}-\frac{1}{4}g_{ab}R)R-\nabla_{a}\nabla_{b}R+g_{ab}\nabla^{2}R\right)-b\nabla^{2}\phi\left(\nabla_{a}\nabla_{b}\phi-\frac{1}{4}g_{ab}\nabla^{2}\phi\right)+\frac{b}{2}(\nabla^{2}\nabla_{a}\phi)\nabla_{b}\phi+\frac{b}{2}(\nabla^{2}\nabla_{b}\phi)\nabla_{a}\phi-\frac{b}{4}g_{ab}(\nabla^{2}\nabla^{c}\phi)\nabla_{c}\phi+\frac{2b}{3}\left((\nabla_{c}\nabla_{a}\phi)(\nabla^{c}\nabla_{b}\phi)-\frac{1}{4}g_{ab}(\nabla_{c}\nabla_{d}\phi)(\nabla^{c}\nabla^{d}\phi)\right)-\frac{b}{3}\nabla^{c}\phi\left(\nabla_{c}\nabla_{a}\nabla_{b}\phi-\frac{1}{4}g_{ab}\nabla_{c}\nabla^{2}\phi\right)-\frac{b}{3}R\left(\nabla_{a}\phi\nabla_{b}\phi-\frac{1}{4}g_{ab}(\nabla\phi)^{2}\right)-\frac{b}{3}\left(R_{ab}-\frac{1}{4}g_{ab}R\right)(\nabla\phi)^{2}+\frac{2b}{3}\left(R_{acbd}-\frac{1}{4}g_{ab}R_{cd}\right)\nabla^{c}\phi\nabla^{d}\phi+\frac{b}{2}(R_{ac}\nabla_{b}\phi+R_{bc}\nabla_{a}\phi)\nabla^{c}\phi-\frac{b}{4}g_{ab}R_{cd}\nabla^{c}\phi\nabla^{d}\phi-\frac{(a+b)}{2\pi}\left(R_{a}{}^{c}R_{bc}-\frac{1}{4}g_{ab}R_{cd}R^{cd}\right)\phi+\frac{(a+b)}{4\pi}\left(R_{adef}R_{b}{}^{def}-\frac{1}{4}g_{ab}R_{cdef}R^{cdef}\right)\phi-\frac{b}{2\pi}\left(R_{acbd}-\frac{1}{4}g_{ab}R_{cd}\right)R^{cd}\phi+\frac{(a+3b)}{12\pi}R\left(R_{ab}-\frac{1}{4}g_{ab}R\right)\phi+\frac{a}{4\pi}\left(\nabla^{2}R_{ab}-\frac{1}{4}g_{ab}\nabla^{2}R\right)\phi-\frac{a}{12\pi}\left(\nabla_{a}\nabla_{b}R-\frac{1}{4}g_{ab}\nabla^{2}R\right)\phi+\frac{(a+b)}{12\pi}\left(\frac{1}{2}\nabla_{a}R\nabla_{b}\phi+\frac{1}{2}\nabla_{b}R\nabla_{a}\phi-\frac{1}{4}g_{ab}\nabla^{c}R\nabla_{c}\phi\right)-\frac{(3a+b)}{12\pi}\left(\frac{1}{2}\nabla_{a}R_{bc}+\frac{1}{2}\nabla_{b}R_{ac}-\frac{1}{4}g_{ab}\nabla_{c}R\right)\nabla^{c}\phi+\frac{(6a+b)}{12\pi}\left(\nabla_{c}R_{ab}-\frac{1}{4}g_{ab}\nabla_{c}R\right)\nabla^{c}\phi-\frac{(a+3b)}{12\pi}R\left(\nabla_{a}\nabla_{v}\phi-\frac{1}{4}g_{ab}\nabla^{2}\phi\right)+\frac{(3a+4b)}{6\pi}\left(R_{acbd}-\frac{1}{4}g_{ab}R_{cd}\right)\nabla^{c}\nabla^{d}\phi-\frac{(3a+7b)}{12\pi}\left(R_{ab}-\frac{1}{4}g_{ab}R\right)\nabla^{2}\phi+\frac{(3a+5b)}{6\pi}\left(\frac{1}{2}R_{ac}\nabla_{b}\nabla^{c}\phi+\frac{1}{2}R_{bc}\nabla_{a}\nabla^{c}\phi-\frac{1}{4}g_{ab}R_{cd}\nabla^{c}\nabla^{d}\phi\right)+\frac{b}{12\pi}\left(\nabla^{2}\nabla_{a}\nabla_{b}\phi-g_{ab}(\nabla^{2}\nabla^{2}\phi+\frac{1}{4}\nabla_{c}R\nabla^{c}\phi-\frac{1}{2}R\nabla^{2}\phi+\frac{3}{2}R_{cd}\nabla^{c}\nabla^{d}\phi)\right)\,,\label{eq:stress-1}
\end{equation}
\end{dmath}that was obtained with the aid of the symbolic algebra
package xAct \citep{Martin-Garcia:2007bqa,Martin-Garcia:2008ysv,Brizuela:2008ra}.
The first term is proportional to the usual Einstein tensor, which
allows us to read off the induced stress tensor via $G_{ab}=8\pi T_{ab}$.
The result can be shown to agree with the semiclassical stress tensor
obtained by Mottola in \citep{Mottola:2016aa}.

The scalar field equation is fourth order in derivatives, but linear
in the $\phi$ field,

\begin{dmath}
\begin{equation}
0=\nabla^{2}\nabla^{2}\phi-\tfrac{2}{3}R\nabla^{2}\phi+2R^{ab}\nabla_{a}\nabla_{a}\phi+\tfrac{1}{3}\nabla^{a}R\nabla_{a}\phi-\frac{(a+b)}{8\pi b}C^{2}-\frac{1}{12\pi}\left(R^{2}-3R_{ab}R^{ab}-\nabla^{2}R\right).\label{eq:scalareom}
\end{equation}
\end{dmath}As a simple check on these expressions, one can take the
trace of the stress tensor in \eqref{eq:stress-1},
\begin{equation}
g^{ab}\left\langle T_{ab}\right\rangle =\frac{b}{2\pi}\left(\nabla^{2}\nabla^{2}\phi-\tfrac{2}{3}R\nabla^{2}\phi+2R^{ab}\nabla_{a}\nabla_{b}\phi+\tfrac{1}{3}\nabla^{a}R\:\nabla_{a}\phi\right)-\frac{1}{16\pi^{2}}\left(c-\frac{2}{3}b\right)\nabla^{2}R\,,\label{eq:stresstrace}
\end{equation}
and eliminate the scalar field using its equation of motion \eqref{eq:scalareom}.
After some straightforward algebra one then recovers the trace anomaly
formula \eqref{eq:traceanom} with $d=e=0$. 

\section{Scalar Field Solution}

We wish to solve the scalar equation of motion \eqref{eq:scalareom}
on the Schwarzschild background,
\begin{equation}
ds^{2}=-\left(1-\frac{2M}{r}\right)dt^{2}+\frac{dr^{2}}{1-\frac{2M}{r}}+r^{2}d\Omega^{2},\label{eq:schwarzschild}
\end{equation}
where $d\Omega^{2}$ is the usual metric on the 2-sphere. Since $R_{ab}=R=0$
in this background, the equation of motion reduces to
\begin{equation}
\nabla^{2}\nabla^{2}\phi=\frac{(a+b)R_{abcd}R^{abcd}}{8\pi b}\,.\label{eq:reducedeom}
\end{equation}
Following \citep{Mottola:2006ew,Mottola:2025fhl}, the general static
spherically symmetric solution can be written as a linear combination
of four independent solutions to the corresponding homogenous problem
plus a special solution $\phi_{P}$ that satisfies the inhomogeneous
equation,
\begin{equation}
\phi=c_{1}\phi_{1}+c_{2}\phi_{2}+c_{3}\phi_{3}+c_{4}+\phi_{P}\,,\label{eq:gensolution}
\end{equation}
where
\begin{align*}
\phi_{1} & =\log\left(1-\frac{2M}{r}\right)\\
\phi_{2} & =r^{2}+4Mr+8M^{2}\log r\\
\phi_{3} & =-\mathrm{Li}_{2}\left(\frac{2M}{r}\right)+\frac{r}{4M}+\frac{3}{2}\log r-\frac{1}{16M^{2}}\left(r^{2}+4Mr-8M^{2}\log r\right)\log\left(1-\frac{2M}{r}\right)\\
\phi_{P} & =-\frac{(a+b)}{8\pi b}\left(\frac{2}{3M}r+2\log r+\frac{1}{12M^{2}}\left(r^{2}+4Mr+8M^{2}\log r\right)\log\left(1-\frac{2M}{r}\right)\right)\,.
\end{align*}
The four constants $c_{i}$ are to be fixed using boundary conditions.

One of the main goals of the present work is to formulate physical
state conditions for this higher-derivative theory. We are interested
in black holes formed via gravitational collapse from smooth initial
data, for which the higher-derivative terms in the field equations
can be treated as small corrections \citep{Lowe:2022tun}. A suitable
initial state could, for instance, consist of an infalling spherical
shell of null matter, surrounding a Minkowski interior, with the initial
shell large enough to ensure weak spacetime curvature on the outside
as well. We expect the higher-derivative theory to be linearly stable
around such a weakly curved spacetime, but we do not expect nonlinear
stability. Indeed, our working hypothesis is that the nonlinear instability
is precisely the Hawking effect, which drives the initial black hole
to shrink in mass, until a Planck mass singular object appears, at
which point the semiclassical equations have broken down. 

Following Hawking's derivation of black hole emission \citep{Hawking1975},
Unruh considered a semiclassical state with vanishing ingoing flux
at past infinity and finite stress tensor on the future horizon \citep{Unruh:1976aa}.
Unruh's ansatz was designed to capture in a simple stationary state
the physics of gravitational collapse into vacuum. His analysis provided
a new insight into Hawking's original calculation, leading to its
wider acceptance. In the present context, we will adopt Unruh-like
conditions on the induced stress tensor of the higher-derivative effective
theory, while keeping in mind some obvious shortcomings. In particular,
without semiclassical back-reaction included, any semiclassical state
that leads to a time-independent outgoing luminosity for the black
hole has an infinite ADM mass. To address this issue, one needs to
consider more general solutions of the time-dependent equations, which
we hope to return to in future work. For the moment, we proceed by
analyzing how physical finiteness conditions on the future horizon
and near past null infinity constrain the parameters of this higher-derivative
gravity solution.

Alternatively, one could consider a broader class of stationary semiclassical
solutions that allow non-vacuum initial data at past infinity. As
shown in \citep{Mottola:2023jlo}, this can lead to large semiclassical
corrections at the future horizon that significantly modify, or even
cancel, the outgoing Hawking flux. This leaves open the possibility
that states with large quantum induced effects are stable, and so
should properly be included as valid initial states. In this case,
it is important to ask whether such end states can also be formed
from smooth initial data. The fact that the analog of the Unruh state
that we derive below is unique, suggests the answer to this question
is no, but a definitive answer can only be obtained by studying the
full time-dependent initial value problem.

\section{Stress Tensor\label{sec:Stress-Tensor}}

Next we evaluate the stress tensor on the Schwarzschild background,
in a semiclassical approximation taking $a,\,b$ and $c$ in \eqref{eq:traceanom}
to be large, so that quantum fluctuations in the scalar $\phi$ and
the gravitational field can be neglected. To find suitable boundary
conditions, we will be mainly interested in the asymptotic form of
the stress tensor as $r\to2M$ and $r\to\infty$. Our goal will be
to impose boundary conditions so that on a Cauchy surface the semiclassical
quantum stress tensor will yield a small correction to the solution
for the metric following the program for dealing with these 4th order
in derivative corrections outlined in \citep{Lowe:2022tun}.

We begin by noticing the stress tensor is independent of $c_{4}$
so this will remain a free parameter, which does not affect observable
quantities. We make the following ansatz for the scalar field
\begin{equation}
\phi(t,r)=d_{p}\,t+\phi(r)\,,\label{eq:scalarsol}
\end{equation}
into the stress tensor, where $\phi(r)$ is a static solution of the
form \eqref{eq:gensolution} and $d_{p}$ is a free parameter to be
later fixed. This simple modification still satisfies the full scalar
equation of motion \eqref{eq:scalareom}. It breaks time-translation
invariance and is a good candidate ansatz for an analog of the Unruh
state, in that the entries in the stress tensor are still time-independent,
but now there is a non-vanishing $T_{rt}$ component. Since we can
ignore $c_{4}$, we are then left with four independent parameters
to be fixed. 

We can derive conditions on the parameters by requiring finite ingoing
null flux across the future horizon. It turns out, the same conditions
come from requiring the $T_{\theta}{}^{\theta}$ component of the
stress tensor to be finite at the future horizon. This amounts to
a simpler calculation and near the horizon we find,
\begin{equation}
T_{\theta}^{\,\,\theta}=\frac{A_{1}}{(r-2M)^{2}}+\frac{A_{1}}{M(r-2M)}+A_{2}\log^{2}\left(\frac{r}{2M}-1\right)+A_{3}\log\left(\frac{r}{2M}-1\right)+\cdots\,,\label{eq:tthetatheta}
\end{equation}
where the $A_{i}$ are quadratic functions of the $c_{i}$. Finiteness
requires setting $A_{1}=A_{2}=A_{3}=0$. As it turns out, this leads
to only two independent constraints on the parameters of \eqref{eq:scalarsol},
which we can use to fix
\begin{align}
c_{3} & =-\frac{a+b}{6\pi b}\nonumber \\
c_{1} & =\frac{(a+b)\log\left(2M\right)}{6\pi b}+2Md_{p}\,.\label{eq:c1c3}
\end{align}
Next we consider the null ingoing flux emerging from past null infinity
$\mathscr{I}^{-}$ obtained by building the null vector $n_{+}^{\mu}=\left(\frac{1}{1-\frac{2M}{r}},1,0,0\right)$
in $(t,r,\theta,\phi)$ coordinates and contracting,
\begin{equation}
n_{+}^{\mu}T_{\mu\nu}n_{+}^{\nu}=B_{1}+\frac{B_{2}}{r}+\frac{B_{3}}{r^{2}}+\cdots\,.\label{eq:nullcontract}
\end{equation}
In order to have a finite energy flux ingoing, we must set $B_{1}=B_{2}=0$
which leads to the single new condition $c_{2}=0$. For these values
of the parameters we then have 
\begin{equation}
B_{3}=\frac{\left(a+b\right)\left(a+b-8\pi M\,b\,d_{p}\right)}{16\pi^{2}bM}\,.\label{eq:b3value}
\end{equation}
To find the analog of the Unruh state, with vanishing ingoing flux
we must set $B_{3}=0$. This determines the value of the remaining
parameter, 
\begin{equation}
d_{p}=\frac{a+b}{8\pi M\,b}\,,\label{eq:dpvalue}
\end{equation}
 and leads to a unique prediction for the outgoing null flux at future
null infinity $\mathscr{I}^{+}$ , with $n_{-}^{\mu}=\left(\frac{1}{1-\frac{2M}{r}},-1,0,0\right)$,
\begin{equation}
n_{-}^{\mu}T_{\mu\nu}n_{-}^{\nu}=\frac{\left(a+b\right)^{2}}{8\pi^{2}M^{2}b\,r^{2}}+\cdots\,,\label{eq:outflux}
\end{equation}
corresponding to an object with finite outgoing luminosity. However,
the unique prediction is unsatisfactory for the following reason.
Conformally coupled matter fields with spin $\leq\nicefrac{1}{2}$
have $b<0$ with $a+b>0$, for which the outgoing energy flux in \eqref{eq:outflux}
is negative! This can be remedied by adding to the gravitational action
a non-local but scale invariant term involving two factors of $C^{2},$
the square of the Weyl tensor. Such a term makes a positive contribution
to the Hawking flux, without affecting the trace anomaly. It may be
localized via the introduction of a second auxiliary scalar field,
as has been considered previously in \citep{Shapiro:1994ww,Balbinot:1999ri}.
The extra scalar field leads to more involved calculations, which
we hope to report on in future work, but our preliminary results indicate
the qualitative conclusion, concerning a uniquely determined Unruh-like
state, remains the same as in the present paper. An alternative, is
to simply stipulate that $b>0$ in the scalar field action \eqref{eq:action}
and view the resulting theory as a toy model that captures key aspects
of the physics of a more realistic two-scalar theory. 

For completeness, the final form for the unique stress tensor is displayed
in the Appendix. Our hope is that having a relatively compact closed-form
expression for the energy flux of Hawking radiation will be useful
for further studies.

\section{Discussion}

We end with a few comments on our results. We have identified a unique
scalar field solution in our model with vanishing ingoing energy flux
but non-vanishing outgoing energy flux. We note that the scaling behavior
of the outgoing flux in \eqref{eq:outflux} satisfies the Stefan-Boltzmann
law for a blackbody in 3+1 dimensions with temperature $T\sim M^{-1}$
and area $A\sim M^{2}$. The energy flux also scales as $N$, the
(large) number of particle species, given our assumption that $a,b=O(N)$. 

The Unruh-like solution \eqref{eq:dpvalue} can be generalized to
a one-parameter family of solutions with finite ingoing flux by not
imposing the final constraint on the $d_{p}$ parameter, but simply
the inequality $B_{3}>0$. However, such solutions violate our condition
that the solution be well-behaved on a Cauchy surface, since they
correspond to static, constant ingoing luminosity, hence infinite
ADM mass. The inadmissible solutions include, in particular, the time-translation
invariant configuration with $d_{p}=0$, which can be viewed as a
Hartle-Hawking-like state with matched ingoing and outgoing energy
fluxes. Such a state would have constant energy density at large $r$
and thus suffers from a Jeans instability.

The main utility of the effective field theory approach advocated
in this paper is that it provides a consistent set of semiclassical
field equations that can in principle be used to solve for time-dependent
semiclassical backgrounds, including evaporating black holes. The
field equations are non-linear and of fourth order in derivatives,
which can lead to unphysical behavior. A promising strategy outlined
in \citep{Lowe:2022tun}, is to work order by order in small $\hbar N$,
starting from a physical initial state, where corrections to the classical
geometry due the induced semiclassical stress tensor are under control
everywhere along the initial time slice. In fact, of the one parameter
family of solutions discussed above, it is only our Unruh-like solution
that can be used as a seed for such a perturbative evaluation. 

There are a number of follow-up directions to pursue. This includes
adding a second auxiliary scalar field in order to obtain positive
outgoing Hawking flux for conventional matter, as discussed above.
Another obvious extension of this work is to evaluate the semiclassical
stress tensor on a time-dependent classical background that describes
gravitational collapse to a black hole, which would allow us to study
the onset of Hawking radiation in this model. Another avenue to consider
is to include electromagnetic fields and discuss Reissner-Nordstrom
black holes. A more ambitious goal is to look for time-dependent solutions
to the full coupled system of scalar field and metric field equations
that describe the formation and subsequent evaporation of a 3+1-dimensional
black hole. 
\begin{acknowledgments}
D.L thanks G. Gabadadze for bringing to his attention \citep{Gabadadze:2023quw},
and J. Donoghue, B.N. Liu and N. Barton for helpful discussions. Work
supported in part by DOE grant de-sc0010010 and the Icelandic Research
Fund grant 228952-053. 
\end{acknowledgments}

\appendix

\section{Closed-form Expression for Stress Tensor}

\begin{dmath*}
\[
T_{tt}=\frac{a+b}{864bM^{2}\ \pi^{2}(2M-r)r^{7}}\Bigl((2M-r)\bigl(8(3821a+4001b)M^{5}-4\ (4655a+4871b)M^{4}r+534(a+b)M^{3}r^{2}+9(43a+47b)M^{2}\ r^{3}+27(a+b)r^{5}\bigr)+8(a+b)M^{2}\log(\frac{2M}{r})\ \bigl(2392M^{4}-2368M^{3}r+556M^{2}r^{2}+12Mr^{3}-3r^{4}+24M^{2}\ (7M^{2}-6Mr+r^{2})\log(\frac{2M}{r})\bigr)\Bigr)
\]
\end{dmath*}\begin{flalign*}
T_{rt}=\frac{(a + b)^2}{32 b M^2 \pi^2 (2 M -  r) r} &&\end{flalign*}\begin{dmath*}
\[
T_{rr}=-\frac{a+b}{864bM^{2}\pi^{2}(2M-r)^{3}r^{5}}\Bigl(3\bigl(16(499a+463b)M^{6}-32(257a+236b)M^{5}r+8(259a+235b)M^{4}r^{2}+72(4a+5b)M^{3}r^{3}-(157a+241b)M^{2}r^{4}-6(a-3b)Mr^{5}+9(a+b)r^{6}\bigr)+8(a+b)M^{2}\log(\frac{2M}{r})\bigl(984M^{4}-848M^{3}r+172M^{2}r^{2}+12Mr^{3}-3r^{4}+8M^{2}(-3M+r)^{2}\log(\frac{2M}{r})\bigr)\Bigr)
\]
\end{dmath*}\begin{dmath*}
\[
T_{\theta\theta}=-\frac{a+b}{864\ bM\pi^{2}r^{4}(-2M+r)^{2}}\Bigl(16(2659a+2371b)M^{5}-32(1445a+1301b)\ M^{4}r+8(1619a+1475b)M^{3}r^{2}+24(23a+29b)M^{2}r^{3}-3\ (143a+191b)Mr^{4}+18(a+3b)r^{5}+8(a+b)M\log(\frac{2\ M}{r})\bigl(1688M^{4}-1608M^{3}r+364M^{2}r^{2}+12Mr^{3}-3r^{4}+\ 8M^{2}(15M^{2}-12Mr+2r^{2})\log(\frac{2M}{r})\bigr)\Bigr)
\]
\end{dmath*}Finally $T_{\phi\phi}=\sin^{2}\theta\,T_{\theta\theta}$.
Other components vanish.

\bibliographystyle{utcaps}
\bibliography{riegert}

\end{document}